\documentstyle[aps,preprint]{revtex}
\input epsf
\global\firstfigfalse  

\tighten

\begin{document}
\pagestyle{empty}

\preprint{
\begin{minipage}[t]{3in}
\end{minipage}
}

\title{Constraints on holographic dark energy from type Ia supernova observations}

\author{Xin Zhang, and Feng-Quan Wu\\\bigskip}
\address{CCAST (World Lab.), P.O.Box 8730, Beijing 100080, People's Republic of
China\\ and\\ Institute of High Energy Physics, Chinese Academy of
Sciences\\ P.O.Box 918(4), Beijing 100049, People's Republic of China\\
}

\maketitle

\begin{abstract}

In this paper, we use the type Ia supernovae data to constrain the
holographic dark energy model proposed by Li. We also apply a
cosmic age test to this analysis. We consider in this paper a
spatially flat Friedmann-Robertson-Walker Universe with matter
component and holographic dark energy component. The fit result
shows that the case $c<1$ ($c=0.21$) is favored, which implies
that the holographic dark energy behaves as a quintom-type dark
energy. Furthermore, we also perform a joint analysis of
SNe+CMB+LSS to this model; the result is well improved, and still
upholds the quintom dark energy conclusion. The best fit results
in our analysis are $c=0.81$, $\Omega_m^0=0.28$, and $h=0.65$,
which lead to the present equation of state of dark energy
$w_0=-1.03$ and the deceleration/acceleration transition redshift
$z_T=0.63$. Finally, an expected SNAP simulation using
$\Lambda$CDM as a fiducial model is performed on this model, and
the result shows that the holographic dark energy model takes on
$c<1$ ($c=0.92$) even though the dark energy is indeed a
cosmological constant.

\vskip .4cm


\vskip .2cm


\end{abstract}

\newpage
\pagestyle{plain} \narrowtext \baselineskip=18pt

\setcounter{footnote}{0}

\section{Introduction}

The type Ia supernova (SN Ia) \cite{sn1,sn2} observations provide
the first evidence for the accelerating expansion of the present
Universe. To explain this accelerated expansion, the Universe at
present time is viewed as being dominated by an exotic component
with large negative pressure referred to as dark energy. A
combined analysis of cosmological observations, in particular, of
the WMAP (Wilkinson Microwave Anisotropy Probe) experiment
\cite{wmap1,wmap2,wmap3}, indicates that dark energy occupies
about 2/3 of the total energy of the Universe, and dark matter
about 1/3. The most obvious theoretical candidate of dark energy
is the cosmological constant $\Lambda$ \cite{cc} which has the
equation of state $w=-1$. An alternative proposal is the dynamical
dark energy \cite{quin,couple} which suggests that the energy form
with negative pressure is provided by a scalar field evolving down
a proper potential. The feature of this class of models is that
the equation of state of dark energy $w$ evolves dynamically
during the expansion of the Universe. However, as is well known,
there are two difficulties arise from all these scenarios, namely
the two dark energy (or cosmological constant) problems --- the
fine-tuning problem and the ``cosmic coincidence'' problem. The
fine-tuning problem asks why the dark energy density today is so
small compared to typical particle scales. The dark energy density
is of order $10^{-47} {\rm GeV}^4$, which appears to require the
introduction of a new mass scale 14 or so orders of magnitude
smaller than the electroweak scale. The second difficulty, the
cosmic coincidence problem, states ``Since the energy densities of
dark energy and dark matter scale so differently during the
expansion of the Universe, why are they nearly equal today''? To
get this coincidence, it appears that their ratio must be set to a
specific, infinitesimal value in the very early Universe.

Recently, considerable interest has been stimulated in explaining
the observed dark energy by the holographic dark energy model. For
an effective field theory in a box of size $L$, with UV cut-off
$\Lambda_c$ the entropy $S$ scales extensively, $S\sim
L^3\Lambda_c^3$. However, the peculiar thermodynamics of black
hole \cite{bh} has led Bekenstein to postulate that the maximum
entropy in a box of volume $L^3$ behaves nonextensively, growing
only as the area of the box, i.e. there is a so-called Bekenstein
entropy bound, $S\leq S_{BH}\equiv\pi M_p^2L^2$. This nonextensive
scaling suggests that quantum field theory breaks down in large
volume. To reconcile this breakdown with the success of local
quantum field theory in describing observed particle
phenomenology, Cohen et al. \cite{cohen} proposed a more
restrictive bound -- the energy bound. They pointed out that in
quantum field theory a short distance (UV) cut-off is related to a
long distance (IR) cut-off due to the limit set by forming a black
hole. In the other words, if the quantum zero-point energy density
$\rho_X$ is relevant to a UV cut-off, the total energy of the
whole system with size $L$ should not exceed the mass of a black
hole of the same size, thus we have $L^3\rho_X\leq LM_p^2$, this
means that the maximum entropy is in order of $S_{BH}^{3/4}$. When
we take the whole Universe into account, the vacuum energy related
to this holographic principle \cite{holoprin} is viewed as dark
energy, usually dubbed holographic dark energy. The largest IR
cut-off $L$ is chosen by saturating the inequality so that we get
the holographic dark energy density
\begin{equation}
\rho_X=3c^2M_p^2L^{-2}~,\label{de}
\end{equation} where $c$ is a numerical constant, and $M_p\equiv 1/\sqrt{8\pi
G}$ is the reduced Planck mass. If we take $L$ as the size of the
current Universe, for instance the Hubble scale $H^{-1}$, then the
dark energy density will be close to the observed data. However,
Hsu \cite{hsu} pointed out that this yields a wrong equation of
state for dark energy. Li \cite{li} subsequently proposed that the
IR cut-off $L$ should be taken as the size of the future event
horizon
\begin{equation}
R_h(a)=a\int_t^\infty{dt'\over a(t')}=a\int_a^\infty{da'\over
Ha'^2}~,\label{eh}
\end{equation} then the problem can be solved nicely and the
holographic dark energy model can thus be constructed
successfully. Some speculations on the deep reasons of the
holographic dark energy were considered by several authors
\cite{holo1}; further studies on this model see also
\cite{holo2,snfit1,snfit2,cmb1,cmb2,cmb3,Nojiri,sfzx,int}. In
addition, it is necessary to discuss about the choice $L=H^{-1}$.
Even though this choice was argued to be unsuitable due to that it
may lead to the holographic dark energy tracks the matter density,
this does not mean that the formalism $M_p^2H^2$ cannot be made
compatible with the observation. There are other contexts in
quantum field theory where one can have a dark energy behaving as
$M_p^2H^2$ without introducing the holographic principle. For
instance, Refs.\cite{QFT} nicely introduce the $M_p^2H^2$ law from
general arguments in quantum field theory. Actually, the first
time where this law was introduced in quantum field theory was in
the context of the renormalization group models of the
cosmological constant; see e.g. \cite{RG}. Otherwise, within the
holographic model framework, the choice $L=H^{-1}$ can also be
favored, if one introduce some interaction between dark energy and
dark matter \cite{int}. However, in this paper we restrict our
attention to the holographic dark energy model proposed by Li
\cite{li}.

In this paper, we will see what constraints to the holographic
dark energy model are set by present and future SNe Ia
observations. Recently, some constraints from SNe Ia on related
model where obtained in Refs.\cite{snfit1,snfit2}. In this paper,
we extend the analysis carried out in Ref.\cite{snfit1}. The work
presented here differs from \cite{snfit1} (and \cite{snfit2}) in
the following aspects: (a) We not only constrain the model by
means of the SNe Ia observations, but also test the fit results by
using the age of the Universe; (b) When performing the analysis of
the SNe data, we also test the sensitivity to the present Hubble
parameter $H_0$ in the fit; (c) For improving the fit result, we
combine the current SNe Ia data with the cosmic microwave
background (CMB) data and the large-scale structure (LSS) data to
analyze the model; (d) We also investigate the predicted
constraints on the model from future SNe Ia observations.

\section{The holographic dark energy model}
The holographic dark energy scenario may provide simultaneously
natural solutions to both dark energy problems as demonstrated in
Ref.\cite{li}. In what follows we will review this model briefly
and then constrain it by the type Ia supernova observations. In
addition, we will also apply a joint analysis of SNe+CMB+LSS data
to this model. Consider now a spatially flat FRW
(Friedmann-Robertson-Walker) Universe with matter component
$\rho_m$ (including both baryon matter and cold dark matter) and
holographic dark energy component $\rho_X$, the Friedmann equation
reads
\begin{equation}
3M_p^2H^2=\rho_m+\rho_{X}~,
\end{equation} or equivalently,
\begin{equation}
{H^2\over H_0^2}=\Omega_m^0a^{-3}+\Omega_X{H^2\over
H_0^2}~.\label{Feq}
\end{equation}
Note that we always assume spatial flatness throughout this paper
as motivated by inflation. Combining the definition of the
holographic dark energy (\ref{de}) and the definition of the
future event horizon (\ref{eh}), we derive
\begin{equation}
\int_a^\infty{d\ln a'\over Ha'}={c\over
Ha\sqrt{\Omega_X}}~.\label{rh}
\end{equation} We notice that the Friedmann
equation implies
\begin{equation}
{1\over Ha}=\sqrt{a(1-\Omega_X)}{1\over
H_0\sqrt{\Omega_m^0}}~.\label{fri}
\end{equation} Substituting (\ref{fri}) into (\ref{rh}), one
obtains the following equation
\begin{equation}
\int_x^\infty e^{x'/2}\sqrt{1-\Omega_X}dx'=c
e^{x/2}\sqrt{{1\over\Omega_X}-1}~,
\end{equation} where $x=\ln a$. Then taking derivative with respect to $x$ in both
sides of the above relation, we get easily the dynamics satisfied
by the dark energy, i.e. the differential equation about the
fractional density of dark energy,
\begin{equation}
\Omega_X '=\Omega_X(1-\Omega_X)(1+{2\over c}\sqrt{\Omega_X})~,
\label{deq}\end{equation} where the prime denotes the derivative
with respect to $x$. This equation describes the behavior of the
holographic dark energy completely, and it can be solved exactly
\cite{li,snfit1},
\begin{equation}
\ln\Omega_X-{c\over 2+c}\ln(1-\sqrt{\Omega_X})+{c\over
2-c}\ln(1+\sqrt{\Omega_X})-{8\over
4-c^2}\ln(c+2\sqrt{\Omega_X})=-\ln(1+z)+y_0~,\label{solution}
\end{equation} where $y_0$ can be determined through
(\ref{solution}) by replacing $\Omega_X$ with
$\Omega_X^0$ as $z=0$. From the energy conservation equation
of the dark energy, the equation of state of the dark energy can
be expressed as
\begin{equation}
w=-1-{1\over 3}{d\ln\rho_X\over d\ln a}~.
\end{equation} Then making use of the formula
$\rho_X={\Omega_X\over 1-\Omega_X}\rho_m^0a^{-3}$ and the
differential equation of $\Omega_X$ (\ref{deq}), the equation of
state for the holographic dark energy can be given
\cite{li,snfit1,snfit2}
\begin{equation}
w=-{1\over 3}(1+{2\over c}\sqrt{\Omega_X})~.\label{w}
\end{equation} We can also give the deceleration parameter $q=-\ddot{a}/aH^2$,
in terms of $\Omega_X$,
\begin{equation}
q={1\over 2}-{1\over 2}\Omega_X-{1\over
c}\Omega_X^{3/2}~.
\end{equation} It can be seen clearly that the equation of state
of the holographic dark energy evolves dynamically and satisfies
$-(1+2/c)/3\leq w\leq -1/3$ due to $0\leq\Omega_X\leq 1$. In this
sense, this model should be attributed to the class of dynamical
dark energy models even though without quintessence scalar field.
The parameter $c$ plays a significant role in this model. If one
takes $c=1$, the behavior of the holographic dark energy will be
more and more like a cosmological constant with the expansion of
the Universe, and the ultimate fate of the Universe will be
entering the de Sitter phase in the far future. As is shown in
Ref.\cite{li}, if one puts the parameter $\Omega_X^0=0.73$ into
(\ref{w}), then a definite prediction of this model, $w_0=-0.903$,
will be given. On the other hand, if $c<1$, the holographic dark
energy will behave like a quintom-type dark energy proposed
recently in Ref.\cite{quintom}, the amazing feature of which is
that the equation of state of dark energy component $w$ crosses
the phantom divide line, $-1$, i.e. it is larger than $-1$ in the
past while less then $-1$ near today. The recent fits to current
SNe Ia data with parametrization of the equation of state of dark
energy find that the quintom-type dark energy is mildly favored
\cite{starob,huterer,running}. Usually the quintom dark energy
model is realized in terms of double scalar fields, one is a
normal scalar field and the other is a phantom-type scalar field
\cite{relev,quintom1}. However, the holographic dark energy in the
case $c<1$ provides us with a more natural realization for the
quintom picture. While, if $c>1$, the equation of state of dark
energy will be always larger than $-1$ such that the Universe
avoids entering the de Sitter phase and the Big Rip phase. Hence,
we see explicitly, the determination of the value of $c$ is a key
point to the feature of the holographic dark energy as well as the
ultimate fate of the Universe.

\vskip.8cm
\begin{figure}
\begin{center}
\leavevmode \epsfbox{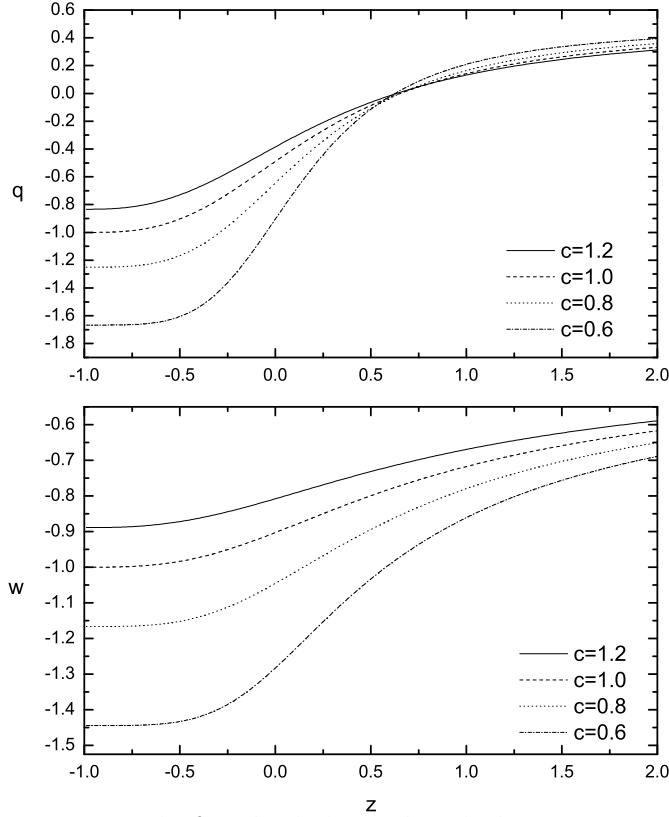} \caption[]{ An illustrative example
for the holographic dark energy model. The evolutions of the
deceleration parameter $q$ and the equation of state of dark
energy $w$. In this case, we take $\Omega_m^0=0.27$.}
\end{center}
\end{figure}

As an illustrative example, we plot in Fig.1 the evolutions of the
deceleration parameter $q$ and the equation of state of dark
energy $w$ in cases of $c=1.2,~1.0,~0.8$ and 0.6, respectively. It
is easy to see that the equation of state of dark energy crosses
$-1$ as $c<1$, which is in accordance with the plots of the
model-independent analysis in Ref.\cite{starob}. It should be
pointed out that the variable cosmological constant model can also
give rise to a quintom behavior, i.e. the effective equation of
state produced by this model can also cross $-1$ \cite{sola}.

In the forthcoming sections, we will see what constraints to the
model described above are set by present and future SNe Ia
observations. In the fitting, we use the recent new high red-shift
supernova observations from the HST/GOODS program and previous
supernova data, and furthermore we test the fit results by means
of the cosmic age data. We find that if we marginalize the
nuisance parameter $h$, the fit of the SNe observation provides
$0.09\lesssim c\lesssim 0.62~(1\sigma)$; this means the
holographic dark energy behaves as a quintom. Further, we find
that the allowed range of model parameters depends on $h$
evidently, for instance, we obtain (in $1\sigma$) $0.21\lesssim
c\lesssim 1.17$ for $h=0.64$, $0.10\lesssim c\lesssim 0.43$ for
$h=0.66$, and $0.03\lesssim c\lesssim 0.07$ for $h=0.71$. The
allowed regions of model parameters become evidently smaller for
larger values of $h$. The fit values of the matter density
$\Omega_m^0$ in this model are apparently larger than the fit
value appears in the $\Lambda$CDM (WMAP result). Now that the SNe
data analysis evidently depends on the value of $h$, an important
thing we should do is to find some observational quantities which
do not depend on $h$ to be useful complements of the SNe data set
to probe the property of holographic dark energy. Such quantities
can be found in CMB and LSS (${\cal R}$ and $A$, respectively, see
section IV). A combined analysis of SNe+CMB+LSS shows that the
confidence region evidently shrinks, and the value of $c$ is
changed considerably, namely in $1\sigma$, $0.65\lesssim c\lesssim
1.04$. We will discuss the cosmological consequences come from the
fits, and compare the case of the joint analysis with the case of
the SNe only in detail. Interestingly, an expected SNAP simulation
using $\Lambda$CDM as a fiducial model shows that the holographic
dark energy model will also favor $c<1$ even though the Universe
is indeed dominated by a cosmological constant.

\section{The current type Ia supernova constraints}

We now perform the best fit analysis on our holographic dark
energy model with data of the type Ia supernova observations. The
luminosity distance of a light source is defined in such a way as
to generalize to an expanding and curved space the inverse-square
law of brightness valid in a static Euclidean space,
\begin{equation}
d_L=\left({{\cal L}\over 4\pi{\cal
F}}\right)^{1/2}=H_0^{-1}(1+z)\int_0^z{dz'\over E(z')}~,
\end{equation} where ${\cal L}$ is the absolute luminosity which
is a known value for the standard candle SNe Ia, $\cal F$ is the
measured flux, $H_0^{-1}$ (here we use the natural unit, namely
the speed of light is defined to be 1) represents the Hubble
distance with value $H_0^{-1}=2997.9h^{-1}$ Mpc, and
$E(z)=H(z)/H_0$ can be obtained from (\ref{Feq}), expressed as
\begin{equation}
E(z)=\left({\Omega_m^0(1+z)^3\over
1-\Omega_X}\right)^{1/2}~,
\end{equation} note that the dynamical behavior of $\Omega_X$ is
determined by (\ref{deq}). The observations directly measure the
apparent magnitude $m$ of a supernova and its red-shift $z$. The
apparent magnitude $m$ is related to the luminosity distance $d_L$
of the supernova through
\begin{equation}
m(z)=M+5\log_{10}(d_L(z)/{\rm Mpc})+25~,
\end{equation} where $M$ is the absolute magnitude which is
believed to be constant for all type Ia supernovae. The numerical
parameter $c$ of the model, the density parameter $\Omega_m^0$ and
the nuisance parameter $h$ can be determined by minimizing
\begin{equation}
\chi^2=\sum_i{[\mu_{\rm obs}(z_i)-\mu_{\rm th}(z_i)]^2\over
\sigma_i^2}~,\label{chisn}
\end{equation} where the extinction-corrected distance moduli
$\mu(z)$ is defined as $\mu(z)=m(z)-M$, and $\sigma_i$ is the
total uncertainty in the observation. The likelihood ${\cal
L}\propto e^{-\chi^2/2}$ if the measurement errors are Gaussian.
In our analysis, we take the 157 gold data points listed in Riess
et al. \cite{Riess} which includes recent new 14 high redshift SNe
(gold) data from the HST/GOODS program. The results of our
analysis for the holographic dark energy model are displayed in
Fig.2. In this figure we marginalize over the nuisance parameter
$h$ and show $68\%$, $95\%$ and $99\%$ confidence level contours,
in the $(c, \Omega_m^0)$-plane. The best fit values for the model
parameters are: $h=0.66$, $\Omega_m^0=0.47^{+0.06}_{-0.15}$, and
$c=0.21^{+0.41}_{-0.12}$ with $\chi_{\rm min}^2=173.44$ (similar
results see also \cite{snfit1}). We see clearly that the fit
values of this model are evidently different from those of
$\Lambda$CDM, i.e. the value of $h$ is slightly smaller and
$\Omega_m^0$ evidently larger (The WMAP results for $\Lambda$CDM
model are \cite{wmap2,PDG}: $h=0.71^{+0.04}_{-0.03}$ and
$\Omega_m^0=0.27\pm 0.04$). We notice in this figure that the
current SNe Ia data do not strongly constrain the parameters
$\Omega_m^0$ and $c$ (in $2\sigma$), in particular $c$, in the
considered ranges. Other observations may impose further
constraints. For instance, the CMB and LSS data can provide us
with useful complements to the SNe data for constraining
cosmological models. But, first, we will apply a cosmic age test
to the SNe analysis.

\vskip.8cm
\begin{figure}
\begin{center}
\leavevmode \epsfbox{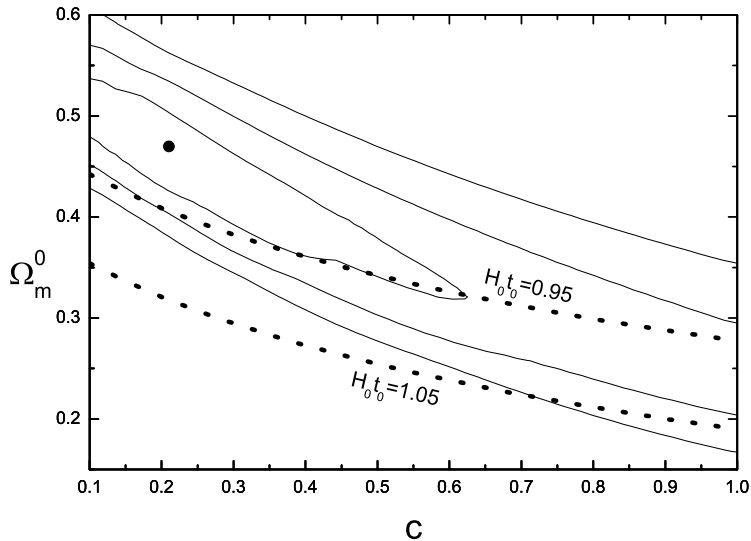} \caption[]{Confidence contours
of $68\%,~95\%$ and $99\%$ in the $(c,\Omega_m^0)$-plane for the
case of marginalizing $h$. The point in the figure, with the
coordinate $(0.21,0.47)$, represents the best fit value, with
$\chi^2_{\rm min}=173.44$. Constraints from the age of the
Universe give $0.95<H_0t_0<1.05$ (at the $1\sigma$ confidence
level), the dashed lines represent these two limits.}
\end{center}
\end{figure}

Recent analyses of the age of old stars \cite{oldstar} indicate
that the expansion time is in the range $11 {\rm Gyr}\lesssim t_0
\lesssim 17{\rm Gyr}$ at $95\%$ confidence, with a central value
$t_0\simeq 13{\rm Gyr}$. Following Krauss and Chaboyer
\cite{oldstar} these numbers add 0.8 Gyr to the star ages, under
the assumption that star formation commenced no earlier than
$z=6$. A naive addition in quadrature to the uncertainty in $H_0$
indicates that the dimensionless age parameter is in the range
$0.72\lesssim H_0t_0\lesssim 1.17$ at $95\%$ confidence, with a
central value $H_0t_0\simeq 0.89$. More recent, Richer et al.
\cite{dwarf1} and Hansen et al. \cite{dwarf2} found an age of
$12.7\pm 0.7$ Gyr at $95\%$ confidence using the white dwarf
cooling sequence method. For a full review of cosmic age see Ref.
\cite{wmap2}. All in all, it seems reasonable to view $\sim 12$
Gyr to be a low limit of the cosmic age \cite{FB}. Now let us
examine the age computation of the holographic dark energy model.
The age of the Universe can be written as
\begin{equation}
t_0=H_0^{-1}\int_0^\infty{dz\over (1+z)E(z)}~,
\end{equation} where $H_0^{-1}$ represents the Hubble time with
value $H_0^{-1}=9.778 h^{-1}$ Gyr. Using the best fit values, the
holographic dark energy model gives the cosmic age $t_0=13.3$ Gyr.
This value is consistent with the above observational analyses.
Furthermore, we impose a more rigorous test on it. The latest
value of cosmic age appears in the Review of Particle Physics
(PDG) \cite{PDG} is given by a combined analysis of various
observations \cite{wmap2}, $t_0=13.7\pm 0.2$. Using the data from
Ref.\cite{PDG}, we can get the dimensionless age parameter range
$0.96\lesssim H_0t_0\lesssim 1.05$ at $68\%$ confidence, with a
central value $H_0t_0\simeq 0.99$. We also display the contours
$H_0t_0=0.96$ and $H_0t_0=1.05$ ($1\sigma$) in Fig.2. It can be
seen explicitly that the $1\sigma$ fit result of the SNe Ia data
is almost excluded by this age test.

\vskip.8cm
\begin{figure}
\begin{center}
\leavevmode \epsfbox{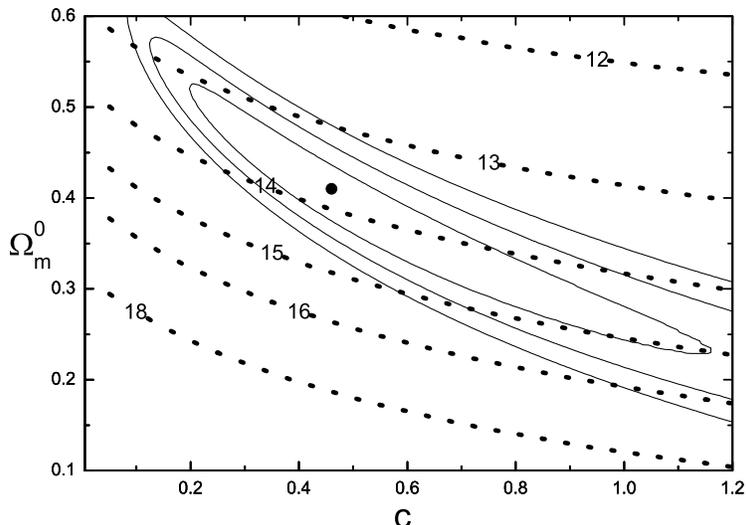} \caption[]{Confidence contours
of $68\%,~95\%$ and $99\%$ in the $(c,\Omega_m^0)$-plane for the
case of $h=0.64$. The best fit values for the parameters are:
$\Omega_m^0=0.41^{+0.12}_{-0.18}$ and $c=0.46^{+0.71}_{-0.25}$,
with $\chi_{\rm min}^2=175.90$. Dashed lines represent the
contours of cosmic age with $t_0=12,~13,~14,~15,~16$ and 18 Gyr.}
\end{center}
\end{figure}

\vskip.8cm
\begin{figure}
\begin{center}
\leavevmode \epsfbox{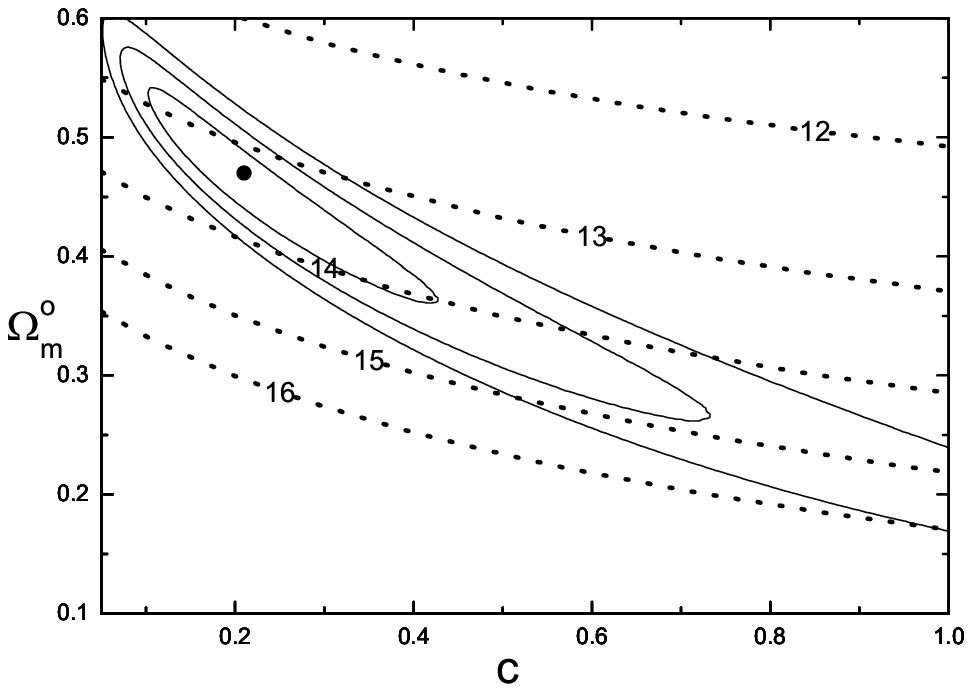} \caption[]{Confidence contours
of $68\%,~95\%$ and $99\%$ in the $(c,\Omega_m^0)$-plane for the
case of $h=0.66$. The best fit values for the parameters are:
$\Omega_m^0=0.47^{+0.07}_{-0.11}$ and $c=0.21^{+0.22}_{-0.11}$,
with $\chi_{\rm min}^2=173.44$. Dashed lines represent the
contours of cosmic age with $t_0=12,~13,~14,~15$ and 16 Gyr.}
\end{center}
\end{figure}

\vskip.8cm
\begin{figure}
\begin{center}
\leavevmode \epsfbox{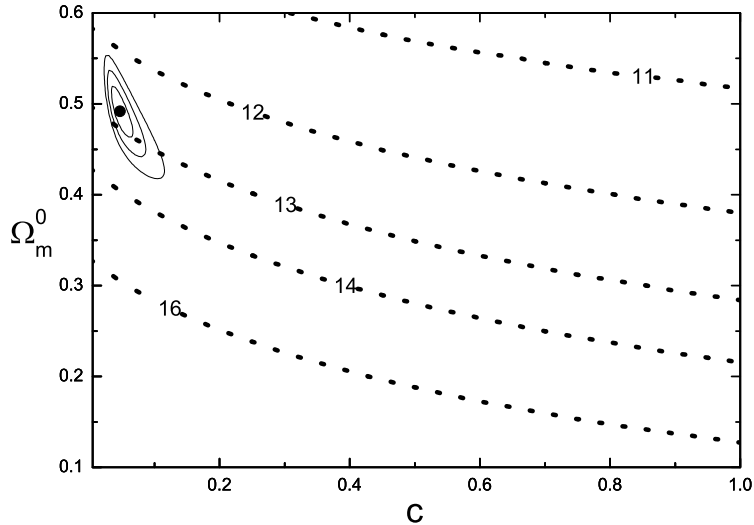} \caption[]{Confidence contours
of $68\%,~95\%$ and $99\%$ in the $(c,\Omega_m^0)$-plane for the
case of $h=0.71$. The best fit values for the parameters are:
$\Omega_m^0=0.49\pm 0.03$ and $c=0.05\pm 0.02$, with $\chi_{\rm
min}^2=175.90$. Dashed lines represent the contours of cosmic age
with $t_0=11,~12,~13,~14$ and 16 Gyr.}
\end{center}
\end{figure}

Next, we will probe the sensitivity to the present Hubble
parameter $H_0$ in the analysis of the SNe data. We check that,
keeping the parameter $h$ fixed, how well the SNe and the cosmic
age will constrain the holographic dark energy model. We fix
$h=0.64,~0.66$ and 0.71, respectively, and show the fit results in
Figs.3-5. From these figures, we notice that with the increase of
the parameter $h$, $c$ decreases evidently and the confidence
contours in the $(c,\Omega_m^0)$-plane shrink sharply. These
figures clearly show that the SNe analysis is dependent on the
parameter $h$ highly. Hence, finding out observational quantities
which do not depend on $H_0$ to jointly constrain the holographic
dark energy model becomes very important and senseful. We also
display age contours in these figures, but the age constraints are
rather weak. The fits of SNe data indicate that the current SNe Ia
data tend to support a holographic dark energy with $c<1$, in
other words, a quintom-type holographic dark energy. However, the
authors of Ref.\cite{snfit2} tried to show another possibility.
They re-examined the holographic dark energy model by considering
the spatial curvature, and found that the holographic dark energy
will not behave as phantom if the Universe is closed. However, the
spatial flatness is a definitive prediction of the inflationary
cosmology, and has been confirmed precisely by the WMAP. Thus we
use the spatial flatness prior throughout this paper.

\section{combined analysis with CMB and LSS}

The above analyses show that the supernovae data alone seem not
sufficient to constrain the holographic dark energy model
strictly. First, the confidence region of $c-\Omega_m^0$ plane is
rather large, especially for the parameter $c$. Moreover, the best
fit value of $\Omega_m^0$ is evidently larger. Second, it is very
hard to understand the fit value of parameter $c$, 0.21, it seems
odd because it leads to an unreasonable present equation of state,
$w_0=-2.64$, the absolute value is too large. Third, even though
the predicted cosmic age $t_0=13.3$ Gyr is larger than the
reasonable low limit of the cosmic age estimated by old stars, a
more rigorous analysis implies that the fit result of the SNe Ia
data is contradictive to the present data of cosmic age
(dimensionless age parameter $H_0t_0$). Furthermore, our analysis
shows that the fit of the SNe Ia data is very sensitive to the
parameter $H_0$. Hence, it is very important to find other
observational quantities irrelevant to $H_0$ as complement to SNe
Ia data. Fortunately, such suitable data can be found in the
probes of CMB and LSS.

In what follows we will perform a combined analysis of SNe Ia,
CMB, and LSS on the constraints of the holographic dark energy
model. We use a $\chi^2$ statistic
\begin{equation}
\chi^2=\chi_{\rm SN}^2+\chi_{\rm CMB}^2+\chi_{\rm LSS}^2~,
\end{equation} where $\chi_{\rm SN}^2$ is given by equation
(\ref{chisn}) for SNe Ia statistics, $\chi_{\rm CMB}^2$ and
$\chi_{\rm LSS}^2$ are contributions from CMB and LSS data,
respectively. For the CMB, we use only the measurement of the CMB
shift parameter \cite{cmbshift},
\begin{equation}
{\cal R}=\sqrt{\Omega_m^0}\int_0^{z_{\rm dec}}{dz\over
E(z)}~,\label{R}
\end{equation} where $z_{\rm dec}=1089$ \cite{wmap1}. Note that this
quantity is irrelevant to the parameter $H_0$ such that provides
robust constraint on the dark energy model. The results from CMB
data correspond to ${\cal R}_0=1.716\pm 0.062$ (given by WMAP,
CBI, ACBAR) \cite{wmap2,cbi}. We include the CMB data in our
analysis by adding $\chi_{\rm CMB}^2=[({\cal R}-{\cal
R}_0)/\sigma_{\cal R}]^2$ (see \cite{wangyun}), where ${\cal R}$
is computed by the holographic dark energy model using equation
(\ref{R}). The only large scale structure information we use is
the parameter $A$ measured by SDSS \cite{sdss}, defined by
\begin{equation}
A=\sqrt{\Omega_m^0}E(z_1)^{-1/3}\left[{1\over
z_1}\int_0^{z_1}{dz\over E(z)}\right]^{2/3}~,\label{A}
\end{equation} where $z_1=0.35$. Also, we find that this quantity
is independent of $H_0$ either, thus can provide another robust
constraint on the model. The SDSS gives the measurement data
\cite{sdss} $A_0=0.469\pm 0.017$. We also include the LSS
constraints in our analysis by adding $\chi_{\rm
LSS}^2=[(A-A_0)/\sigma_A]^2$ (see \cite{gyg}), where $A$ is
computed by the holographic dark energy model using equation
(\ref{A}).

Note that we have chosen to use only the most conservative and
robust information, ${\cal R}$ and $A$, from CMB and LSS
observations. These measurements we use do not depend on the
Hubble parameter $H_0$. Furthermore, by limiting the amount of
information that we use from CMB and LSS observations to
complement the SNe Ia data, we minimize the effect of the
systematics inherent in the CMB and LSS data on our results.
Figure 6 shows our main results, the contours of $1\sigma$,
$2\sigma$, and $3\sigma$ confidence levels in the $c-\Omega_m^0$
plane. The best fit values for the model parameters are: $h=0.65$,
$\Omega_m^0=0.28\pm 0.03$, and $c=0.81^{+0.23}_{-0.16}$, with
$\chi_{\rm min}^2=176.67$. We see clearly that a great progress
has been made when we perform a joint analysis of SNe Ia, CMB, and
LSS data. Note that the best fit value of c is also less than 1,
while in $1\sigma$ range it can slightly larger than 1. Now the
fit value of $\Omega_m^0$ is roughly as the same as that of the
$\Lambda$CDM model (WMAP result), but $h$ is still slightly
smaller. It should be pointed out that a slightly lower value of
$h$ is, however, in agreement with the observations of \cite{hobs}
which can accommodate lower values of $h\sim 0.6$. We also see
from the figure that the fit result of the joint analysis is
consistent with the dimensionless age range $0.96\lesssim
H_0t_0\lesssim 1.05$.

\vskip.8cm
\begin{figure}
\begin{center}
\leavevmode \epsfbox{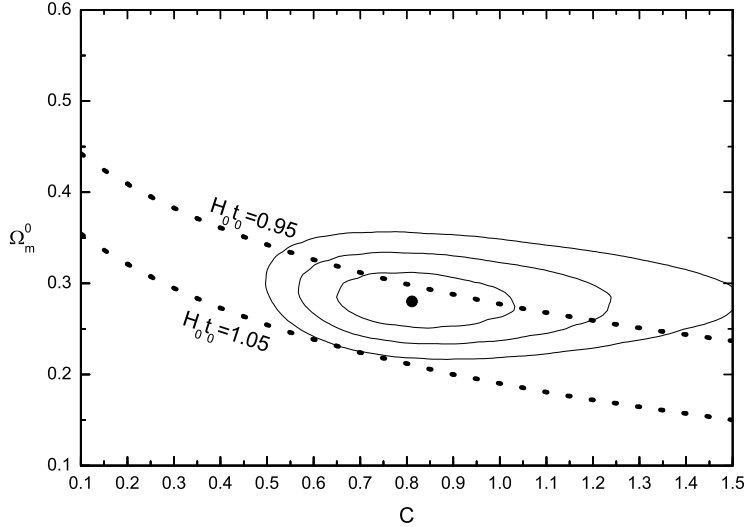} \caption[]{Results for
analysis of SNe+CMB+LSS data, $1\sigma$, $2\sigma$, $3\sigma$
confidence levels in the $c-\Omega_m^0$ space, marginalizing over
$h$. The best fit values are: $h=0.65$, $\Omega_m^0=0.28$, and
$c=0.81$, with $\chi_{\rm min}^2=176.67$. Cosmic age test, the
dimensionless age parameter range $0.96\lesssim H_0t_0\lesssim
1.05$, also shown.}
\end{center}
\end{figure}

We now compare the fit results of the SNe analysis and the joint
analysis of SNe, CMB, and LSS, and discuss the different
cosmological consequences. Let us first look at the likelihood
distributions of the parameter $c$ in the two fits. In Fig.7 we
plot the 1-dimensional likelihood function for $c$, marginalizing
over the other parameters. The big difference can lead to rather
different conclusions for some important cosmological parameters,
such as today's equation-of-state parameter of dark energy,
today's deceleration parameter, etc. For the best fit results of
the two fits, we plot the evolution behaviors of the deceleration
parameter $q$ and the equation-of-state parameter of dark energy
$w$ in Fig.8. For the SNe+CMB+LSS joint analysis, from the figure,
we see that the deceleration parameter $q$ has a value of
$q_0=-0.61$ at present. The transition from deceleration to
acceleration ($q(z_T)=0$) occurs at a redshift of $z_T=0.63$. The
equation-of-state parameter $w$ is slightly smaller than $-1$ at
present, $w_0=-1.03$. For comparison, we list these values for the
alone analysis of SNe data: $q_0=-1.60$, $z_T=0.27$, and
$w_0=-2.64$. Therefore, obviously, the results of the combined fit
seem more reasonable. From a joint analysis of SNe+CMB+LSS data,
one may obtain within the framework of the holographic dark energy
model a fairly good idea of when the Universe began to accelerate
and how fast the present acceleration is. By contraries, the
cosmological consequences given by the alone analysis of the SNe
data seem unreasonable. Comparing our plots in Fig.8 with the
model-independent plots in Ref.\cite{starob} (which also use data
from \cite{Riess}), we find that the holographic plot for $c=0.81$
case is in good agreement with those model-independent plots for
the redshift range $z=0-2$, while the $c=0.21$ case dose not
accord. Moreover, it should be mentioned again that these results
demonstrate that the best fit to SNe+CMB+LSS observations also
favors a quintom-type holographic dark energy.

\vskip.8cm
\begin{figure}
\begin{center}
\leavevmode \epsfbox{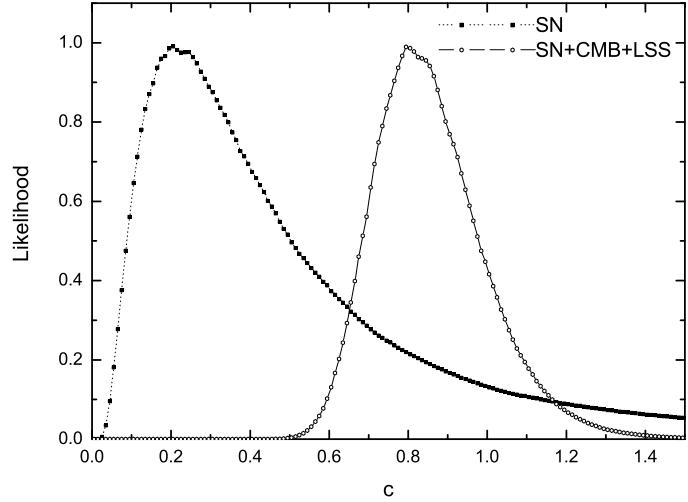} \caption[]{Likelihood
distributions of parameter $c$ in the fits of SNe only and
SNe+CMB+LSS.}
\end{center}
\end{figure}

\vskip.8cm
\begin{figure}
\begin{center}
\leavevmode \epsfbox{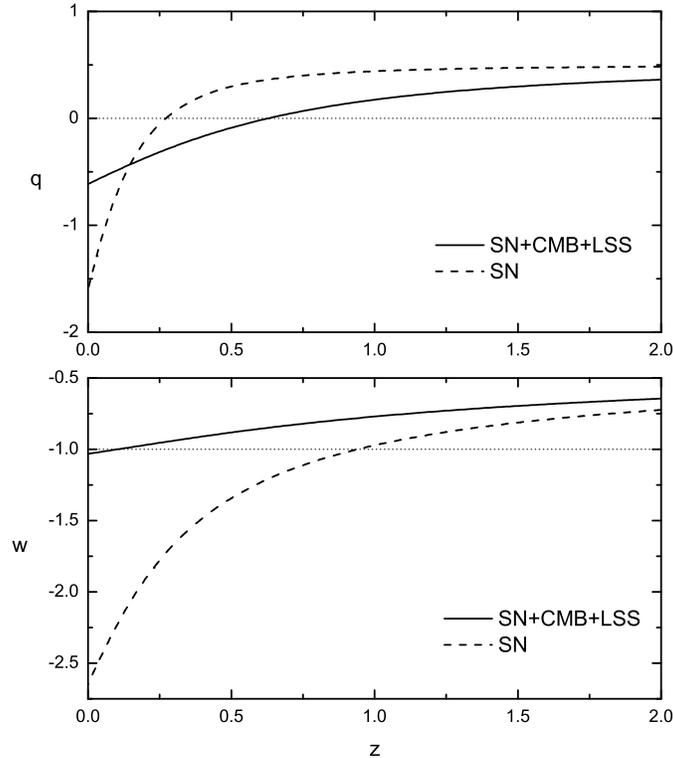} \caption[]{Deceleration parameter
$q$ and equation of state of dark energy $w$, versus red-shift
$z$, from two fits, SNe only and SNe+CMB+LSS. Solid lines
correspond to the joint analysis of SNe+CMB+LSS, with the
parameters $c=0.81$, $\Omega_m^0=0.28$, and $h=0.65$. Dashed lines
correspond to the alone analysis of SNe, with the parameters
$c=0.21$, $\Omega_m^0=0.47$, and $h=0.66$.}
\end{center}
\end{figure}

\section{Expected SNAP analysis}

Finally we consider how well the proposed Supernova/Acceleration
Probe (SNAP), may constrain the parameters $c$ and $\Omega_m^0$.
The SNAP mission is expected to observe about 2000 type Ia SNe
each year, over a period of three years, according to the SNAP
specifications.To find the expected precision of the SNAP, one
must assume a fiducial model, and then simulate the experiment
assuming it as a reference model. We will use SNAP specifications
to construct mock SNe catalogues \cite{snap}. Following previous
investigations \cite{sne}, we assume, in our Monte Carlo
simulations, that a total of 2000 supernovae (roughly one year of
SNAP observations) will be observed with the following redshift
distribution. We consider, 1920 SNe Ia, distributed in 24 bins,
from $z=0$ to $z=1.2$. From redshift $z=1.2$ to $z=1.5$, we assume
that 60 SNe Ia will be observed and we divide them in 6 bins. From
$z=1.5$ to $z=1.7$ we consider 4 bins with 5 SNe Ia in each bin.
All the supernovae are assumed to be uniformly distributed with
$\Delta z=0.05$. To fully determine the $\chi^2$ functions, the
error estimates for SNAP must be defined. Following \cite{snap},
we assume that the systematic errors for the apparent magnitude,
$m$, are given by

\begin{equation}
\sigma_{sys}={0.02 \over1.5}z ~~,
\end{equation} which are measured in
magnitudes such that at $z=1.5$ the systematic error is $0.02$
mag, while the statistical errors for $m$ are estimated to be
$\sigma_{sta}=0.15$ mag. We add both kinds of errors quadratically

\begin{equation}
\sigma_{mag}(z_i)=\sqrt{\sigma^2_{sys}(z_i)+{\sigma^2_{sta} \over
n_i}}~~, \end{equation} where $n_i$ is the number of supernovae in
the $i'th$ redshift bin with width $\Delta z\approx0.05$. Now let
us assume a spatial flat $\Lambda$CDM model as a fiducial model,
and analyze the holographic dark energy model fit. We aim to show
if the Universe is indeed described by the $\Lambda$CDM model, how
well the the fitting of a holographic dark energy model to be.

\vskip.8cm
\begin{figure}
\begin{center}
\leavevmode \epsfbox{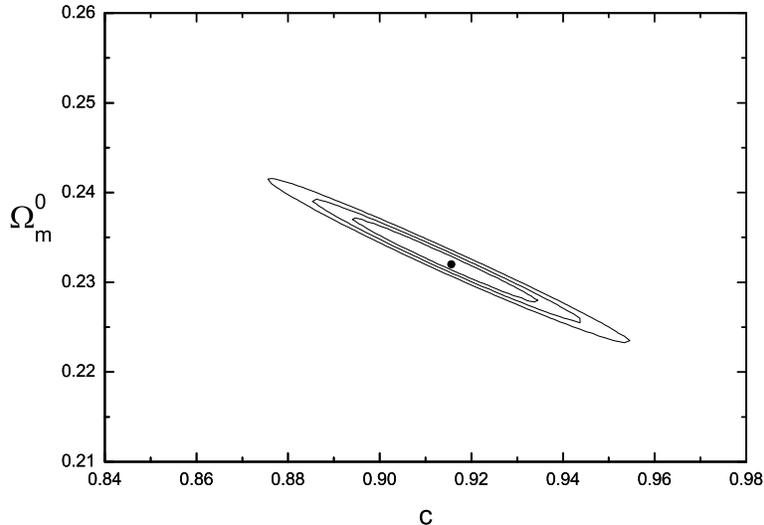} \caption[]{Predicted confidence
contours of $68\%,~95\%$ and $99\%$ in the $(c,\Omega_m^0)$-plane
for the SNAP mission are shown. We consider a fiducial model with
$\Omega_m^0=0.27$ and $h=0.71$. The best fit values for the
parameters are: $\Omega_m^0=0.23$ and $c=0.92$.}
\end{center}
\end{figure}

In Fig.9 we display the results of our simulation assuming a
$\Lambda$CDM model as fiducial model with $\Omega_m^0=0.27$ and
$h=0.71$. In the fit, we marginalize over the nuisance parameter
$h$. The best fit values for the model parameters are:
$\Omega_m^0=0.23$ and $c=0.92$. From this figure it is clear that
SNAP mission is able to place rigorous constraints on the
holographic dark energy model. On the other hand, we notice with
interest that, even though the dark energy in the Universe is
exactly described by a cosmological constant $\Lambda$, the
precision type Ia supernova observations will still support a
quintom-type holographic dark energy.

\section{Concluding remarks}

In this paper, we investigated constraints on the holographic dark
energy model from current and future SN Ia observations. We
considered a spatially flat FRW Universe with matter component and
holographic dark energy component. For the holographic dark energy
model, the numerical constant $c$ plays a very important role in
determining the evolutionary behavior of the space-time as well as
the ultimate fate of the Universe. In a holographic dark energy
dominated Universe, the case of $c=1$ corresponds to an asymptotic
de Sitter Universe; The choice of $c<1$ will lead to dark energy
behaving as quintom, and in this case, the Gibbons-Hawking entropy
will eventually decrease as the event horizon shrink such that
violates the second law of thermodynamics; While the case of $c>1$
does not violate the second law of thermodynamics, in this
situation, the evolution of the corresponding space-time avoids
entering de Sitter phase and Big Rip phase. Though the choice of
$c=1$ is favored theoretically, other possibilities can not be
ruled out from the viewpoint of phenomenology, and only
experiments and observations are capable of determining which
choice is realistic. We derived model parameter ranges from the
analysis of the present available SNe Ia data, and then imposed a
rigorous test using the new data of age of the Universe on the
derived parameter region. However, only the supernova analysis
seems not sufficient to be able to precisely determine the value
of $c$. For improving the result of the analysis, we perform a
joint analysis of SNe, CMB, and LSS data to the holographic dark
energy model.

The results of the SNe analysis show that the holographic dark
energy model behaves as quintom in $1\sigma$ confidence level,
consistent with the current SNe Ia data. However, when we perform
a rigorous age test on this analysis result using the latest PDG
data, the $1\sigma$ allowed range is almost ruled out. Moreover,
in this case, the confidence regions in the parameter-plane are
rather large. We also probe the sensitivity to the present Hubble
parameter $H_0$ in this analysis. We do this by fixing the
parameter $h$, and find the allowed regions of the parameters
shrink sharply as $h$ increases. Hence, a combined analysis of SNe
data with other observational data becomes important. A joint
analysis of SNe+CMB+LSS produces more reasonable results:
$c=0.81$, $\Omega_m^0=0.28$, and $h=0.65$, leading to the present
equation of state of dark energy is $w_0=-1.03$, and the epoch at
which the Universe began to accelerate is $z_T=0.63$. The
confidence regions in this analysis become more compact. On the
whole, the analysis indicates that the case of $c<1$ is consistent
with the present SNe Ia data.

We are also interested in the constraints on the holographic dark
energy model from a fictitious future supernova experiment. An
expected SNAP fit, by using the $\Lambda$CDM model as fiducial
model to generate mock observational data, shows that the case
$c<1$ ($c=0.92$) also favored. Obviously, a large number of
supernovae at high redshifts, as well as better knowledge of the
values of $H_0$ and $\Omega_m^0$ are therefore required to draw
firm conclusions about the property of the holographic dark
energy. We expect that a more sophisticated combined analysis of
various observations will be capable of determining the value of
$c$ exactly and thus revealing the property of the holographic
dark energy.

\begin{acknowledgments}
We would like to thank Orfeu Bertolami, Zhe Chang, Ling-Mei Cheng,
Bo Feng, Yungui Gong, Alan Heavens, Qing-Guo Huang, Hong Li, Miao
Li, Ramon Miquel, Junqing Xia, and Xinmin Zhang for helpful
discussions. This work was supported by the Natural Science
Foundation of China (Grant No. 10375072).
\end{acknowledgments}



\end{document}